\pdfoutput=1
\documentclass[sigconf]{acmart}
\AtBeginDocument{%
  \providecommand\BibTeX{{%
    \normalfont B\kern-0.5em{\scshape i\kern-0.25em b}\kern-0.8em\TeX}}}

\usepackage{lineno}
\usepackage{amsmath}
\usepackage{multirow}
\usepackage{enumitem}
\usepackage{url}
\usepackage{stfloats}
\usepackage{scalefnt}
\usepackage{enumitem}

\copyrightyear{2023}
\acmYear{2023}
\setcopyright{none}\acmConference[SBES 2023]{\textbf{English Version} of the Paper published as part of the Proceedings of the XXXVII Brazilian Symposium on Software Engineering}{September 25--29, 2023}{Campo Grande, Brazil}
\acmBooktitle{\textbf{English Version} of the Paper published as part of the Proceedings of the XXXVII Brazilian Symposium on Software Engineering (SBES 2023), September 25--29, 2023, Campo Grande, Brazil}
\acmPrice{15.00}

\begin{document}

\title{Gamification in Software Engineering Education: \\a Tertiary Study [English Version]}

\author{Simone de França Tonhão}
\affiliation{%
  \institution{Universidade Estadual de Maringá}
  \city{Maringá, PR, Brazil }
  \country{} 
}
\email{siimone.franca@gmail.com}

\author{Júlio Herculani}
\affiliation{%
  \institution{Universidade Estadual de Maringá}
  \city{Maringá, PR, Brazil }
  \country{}
}
\email{juliobudiskiherculani@gmail.com}

\author{Marcelo Y. Shigenaga}
\affiliation{%
  \institution{Universidade Estadual de Maringá}
  \city{Maringá, PR, Brazil }
  \country{}
}
\email{marcelo.shigenaga@gmail.com}

\author{Andressa S. S. Medeiros}
\affiliation{%
  \institution{Universidade Estadual de Maringá}
  \city{Maringá, PR, Brazil }
  \country{}
}
\email{pg404056@uem.br}

\author{Aline M. M. M. Amaral}
\affiliation{%
  \institution{Universidade Estadual de Maringá}
  \city{Maringá, PR, Brazil }
  \country{}
}
\email{ammmamaral@uem.br}

\author{Williamson Silva}
\affiliation{%
  \institution{Universidade Federal do Pampa}
  \city{Alegrete, RS, Brazil}
  \country{}
}
\email{williamsonsilva@unipampa.edu.br}

\author{Thelma E. Colanzi}
\affiliation{%
  \institution{Universidade Estadual de Maringá}
  \city{Maringá, PR, Brazil }
  \country{}
}
\email{thelma@din.uem.br}

\author{Igor F. Steinmacher}
\affiliation{%
  \institution{Northern Arizona University}
  \city{Flagstaff, AZ, USA}
  \country{}
}
\email{igor.steinmacher@nau.edu}

\renewcommand{\shortauthors}{Simone de França Tonhão, et al.}
\renewcommand{\shorttitle}{Gamification in Software Engineering Education: a Tertiary Study}

\renewcommand{\abstractname}{Abstract}
\begin{abstract}



As the significance of Software Engineering (SE) professionals continues to grow in the industry, the adoption of gamification techniques for training purposes has gained traction due to its potential to enhance class appeal through game-derived elements.
This paper presents a tertiary study investigating the application of gamification in Software Engineering (SE) education. The study was conducted in response to recent systematic literature reviews and mappings on the topic. The findings reveal that the areas of SE most frequently gamified are Software Testing and Software Quality, with competition and cooperation being the most commonly utilized gamification elements. Additionally, the majority of studies focus on structural gamification, where game elements are employed to modify the learning environment without altering the content. The results demonstrate the potential of gamification to improve students' engagement and motivation throughout the SE learning process, while also impacting other aspects such as performance improvement, skill development, and fostering good SE practices. However, caution is advised as unplanned and incorrectly applied gamification measures may lead to significant declines in performance and motivation.

\end{abstract}


\maketitle


\section{Introduction}
\label{sec:introduction}

Several changes have occurred in Software Engineering (SE) over time, such as the profession which was initially lab-oriented and later shifted to a more industry-oriented process \cite{wang2000software}. Traditionally, SE focused on programming methodologies, 
software development models, and development tools \cite{wang2000software}. It is notable that in the evolution of the field, many changes occur, not only in the profession but also in the competencies desired by the industry. This requires a mastery of various concepts in the field by recent graduates, who may not be prepared for the "practical" demands of the market.

This gap exists because many innovations originate from the industry, and academia must balance basic and social skills with realistic projects \cite{8808915}. The profession of software engineer and education in Software Engineering (SE) are challenging, as they demand many different qualitative attributes. Skills such as divergent thinking and collaborative learning are important for the practice of software engineers \cite{6401119}.

SE education is still mostly conducted in a traditional manner, as pointed out by \citet{8658908}, using deductive or expository methodologies. Generally, using small and fictitious examples, which creates a distance between students and the content taught and the industrial projects \cite{kapp2012gamification,8658908}.

There are also issues related to the labor market that directly affect SE education, such as the need for professionals with personal (\textit{soft skills}), leadership, and technical skills~\cite{5958974}. Since its inception, SE has been improving, and education must keep up with its changes. This education currently faces one of the biggest challenges, which is student engagement \cite{9125353}.

One alternative to make education more practical and engage students is gamification. One of the proposals of gamification is to make classes more attractive through the dynamism offered by games. This is a concept that has been gaining ground in SE education, especially when related to motivation and engagement. Gamification involves using, in other contexts, the elements traditionally found in games, such as narrative, feedback system, reward system, fun, competition, trial and error, among others \citep{Fardo2013}.

These elements are used with the intention of achieving the same engagement and motivation that can usually be observed in players who interact with highly successful \textit{games}. The elements found in games act as a mechanism of motivation for the individual, contributing to their involvement in various environments \citep{PintoSilva2017}. In this context, gamification can be applied in various fields of human activity, since the language and methodology of \textit{games} can be effective in solving problems in various areas \citep{Fardo2013}.

Gamification used in conjunction with an active methodology supports the teaching process, providing engagement during the learning experience \citep{PintoSilva2017}. Thus, through an active methodology that gives students autonomy to explore their knowledge, they become protagonists of learning that using gamification, becomes motivating and can encourage studies \citep{FigasEtAl2013}.

Therefore, the goal of this study was to investigate the use of gamification in the context of SE education through a tertiary study. The study was guided by four research questions, which focused on the following aspects: (RQ1) how gamification has been applied, (RQ2) in which areas of SE, (RQ3) which elements were used, and (RQ4) what are the impacts (positive or negative) of using such an approach in SE education. A tertiary study was chosen, since there were already systematic reviews and systematic mappings of the literature relatively recent~\cite{alhammad2018, jesus2018, indriasari2020, machuca2018, darejeh2016} in the major area of interest of this study. This situation motivated the execution of the tertiary aiming to perform a meta-analysis around the use of gamification in SE education.

This study followed the systematic review protocol defined by \citet{kitchenham2004}, divided into three basic phases, which are: planning, execution (conducting, study selection, information extraction), and analysis of the results. Additionally, the StArt tool \cite{start}, was used to support the execution of the processes.

Our results showed that gamification has the potential to increase the engagement and motivation of students in SE education, also acting on other points such as improvement in performance; teamwork and leadership; and stimulation of good SE practices. However, gamification can be contradictory when used in teaching, as its application in an incorrect manner, or through the use of elements not suitable

\section{Gamification}
\label{sec:gamification}

Gamification originated from games, which can be defined in various ways. Among them, \citet{koster250theory} define games as a "system that exposes the player to abstract challenges governed by a set of rules, interactivity, and continuous feedback, with predetermined outcomes that evoke an emotional reaction." Additionally, \citet{salen2004rules} defines games as a "system that engages the player in a virtual conflict governed by a set of rules containing a known and justified outcome," while \citet{mcgonigal2011reality} states that "a game has a goal that provides a purpose, rules that are the limitations to achieve the goal, feedback that evaluates the player and how close they are to the goal, and voluntary participation accepting the goal, rules, and feedback provided by the game."

By utilizing the elements that compose games, it is possible to assist in the education of students who are familiar with these resources and technologies, which is determined as learning gamification. In this context, the most recognized definition of Gamification by academia is formed by \citet{kapp2012gamification}, as the use of game-based mechanics, aesthetics, and logic to engage people, motivate actions, promote learning, and solve problems. \citet{deterding2011gamification} also define gamification as the use of game elements outside of their original context.

According to \citet{kapp2012gamification}, gamification requires components allied with a rule to provoke interaction and emotional states that games provide through aesthetics. In addition to aesthetics, in the mechanics are the elements used in games, such as leaderboards, point systems, difficulty levels, time constraints, and badges.

It is worth noting that serious games, often related to gamification, are a different experience and designed using game mechanics and game thinking to educate individuals in a specific content domain, unlike gamification. Serious games use the concept of gamification as a trivial use of game mechanics to artificially engage students and others in activities they would not otherwise engage in. Examples of serious games include leaderboards and artificially inflated high scores in real-life situations \cite{kapp2012gamification}.

\citet{toda2019} grouped game elements into five dimensions: fictitious, performance, ecological, social, and personal. Each dimension contains specific game elements, such as levels, points, and progression, which have a performance character in the game.
These game elements together form the components that, when combined with a rule, provoke interaction and provide the experience that a game offers through appearance. Thus, the use of elements can be related to the interest in designing a specific state in the student, thus choosing the appropriate elements to generate the desired interaction.

Among the various elements and possibilities involving the implementation of gamification in education, there are two predominant strategies: structural gamification and content gamification~\cite{kapp2013gamification}. Structural gamification uses game elements in the learning process to motivate students to participate in suggested activities through extrinsic motivation~\cite{kapp2013gamification}. In this case, the structure housing the content is gamified, while the content itself undergoes no modification, remaining unchanged \cite{kapp2013gamification}. Applying elements such as badges, points, and leaderboards in the medium used for teaching, for example, constitutes structural gamification~\cite{kapp2012gamification}. The teaching medium can be an online platform and the classroom, which can also receive elements in a creative and non-virtual way. It is up to the teacher, instructor, to find the best way to apply the elements in their own context, with the aim of satisfying their needs to support teaching and learning.

In content gamification, part or all of the theoretical study material of a task, course, or discipline is modified~\cite{kapp2013gamification}. Thus, greater interaction between participants and leadership by students is sought through voluntary and intentional participation in the learning process. Using elements such as narrative that brings the student closer to the material used during learning, making them the protagonist of the learning experience, constitutes content gamification \citep{kapp2012gamification}. The elements are applied directly to the content material such as in textbooks, slides, and video lessons. Therefore, the student is the protagonist of a story in which they are the main character, resulting in exposure to the contents and learning during the performance of the tasks of the gamified activity.

As a way to visualize the potential of gamification in education, one can observe some of its uses for learning. \url{Codecombat.com} is used by students to learn programming concepts and is available in 50 languages. Since 2013, it has captivated around 5 million players in over 200 countries, totaling more than 1 billion lines of code through many different languages. During interactions with action characters in an engaging narrative, the student can learn programming concepts and apply them in practice. In this example, both structural gamification and content gamification were used, as the platform uses elements such as scoring and the content material contains elements such as the narrative, which presents a character controlled by the student and whose progress is through the course of a story.

We can also see the use of effective gamification in the non-profit organization called Code.org, which has tens of millions of accesses to its gamified programming activities available in 67 languages and accessed in over 180 countries. Containing a variety of practical activities for children, young people, and adults, the platform has already been used in more than 15,000 schools, helping to understand computational logic and Computer Science through game elements such as the progress bar and narratives. We can observe that in this example, both structural and content gamification were used, as it uses narratives in the learning content and progress bars in the learning environment.

Finally, we can mention Duolingo, one of the most famous and acclaimed language learning platforms with over 300 million students, with various awards including Google's Best of the Best in two consecutive years and valued at \$1.5 billion, which refers to another example of a successful use of structural gamification, with elements such as points, progress bars, and leaderboards, and content gamification, with elements such as narrative and storytelling. This platform is completely free and available as a mobile app for Android, iOS, Windows Phone, and web availability.

In addition, various online teaching platforms also use gamification with game elements to engage students. Among them, those with some of the highest usage and search rates on Google Trends are highlighted: Moodle, Coursera, and Udemy.
\section{Research Method}
\label{sec:methodology}

\begin{figure}[!t]
\centering
\includegraphics[width=1.0\linewidth]{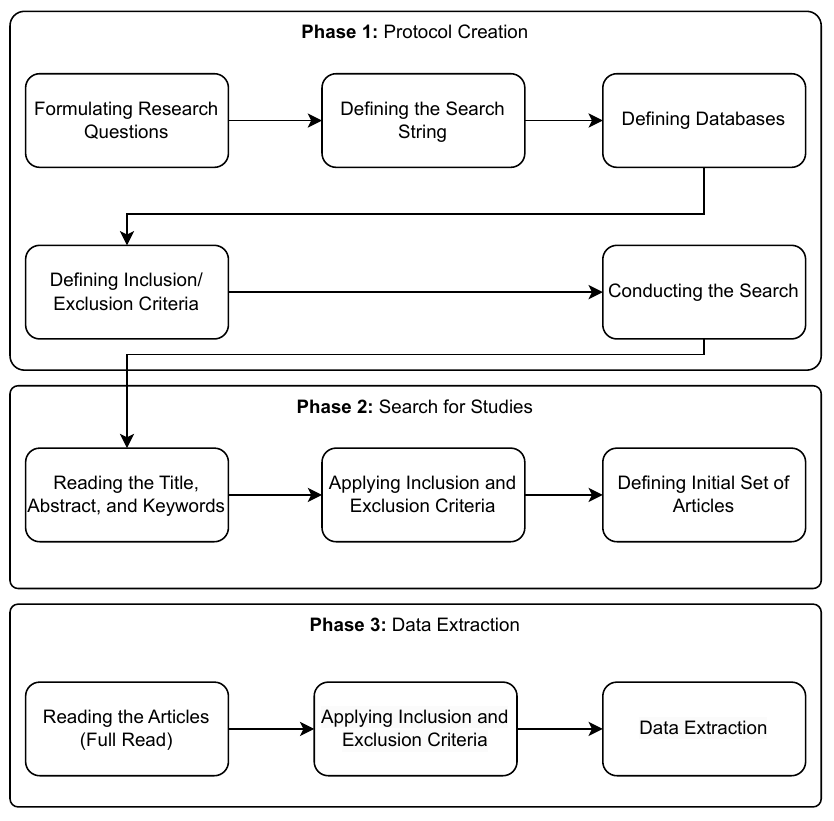}
\caption{Steps of the study methodology.}
\label{fig:process}
\end{figure}

In this section, we describe the method followed for conducting this tertiary study. The planning of the study protocol was carried out following the basic phases of systematic reviews defined by \citet{kitchenham2004}. The execution and data extraction were performed using the StArt tool, and the generated file is available in a repository for future use~\cite{repositorio}. Figure~\ref{fig:process} presents the three defined phases of the study and their respective activities.

The first step of \textbf{Phase 1} of the study was the drafting of the protocol, in which the study's objectives, research questions, keywords, search string, databases, inclusion and exclusion criteria, and the steps to be followed during the process were defined. The selected publications were secondary studies such as systematic mappings or systematic literature reviews.

The main objective of this study is an overview of the use of gamification in SE education. Gamification in education is an emerging topic, especially with changes introduced by the pandemic. However, it is noted that knowledge remains scattered without a source that aggregates the results concerning the elements and their impacts. Thus, the research questions of this study were formulated to provide insights into the potential use of gamification in SE education. It is important to understand how gamification has been used, in which areas of SE, which elements are most used, and how the use of this approach can impact teaching.

Following this, the \textbf{research questions were formulated}. These questions are presented in Table~\ref{tab:questions}, along with their respective objectives. The process of search, selection, and data extraction will be presented in detail in the following sections.

\begin{table}[!t]
\caption{Research questions and their objectives.}
\label{tab:questions}
\centering
\footnotesize
\begin{tabular}{l|l}
\hline
\textbf{Research Question}& \textbf{Objective} \\ \hline
\begin{tabular}[c]{@{}l@{}}RQ1. What strategies are used for applying\\ gamification in SE education?\end{tabular} & \begin{tabular}[c]{@{}l@{}}Identify and distinguish\\ gamification strategies.\end{tabular} \\ \hline
\begin{tabular}[c]{@{}l@{}}RQ2. Which SE knowledge areas have been\\ gamified?\end{tabular}       & \begin{tabular}[c]{@{}l@{}}Identify and classify\\ the knowledge areas.\end{tabular}     \\ \hline
\begin{tabular}[c]{@{}l@{}}RQ3. Which gamification elements are used?\end{tabular}                                   & \begin{tabular}[c]{@{}l@{}}Identify, categorize\\ and classify the elements.\end{tabular}    \\ \hline
\begin{tabular}[c]{@{}l@{}}RQ4. What are the impacts of gamification in SE\\ education?\end{tabular}                                & \begin{tabular}[c]{@{}l@{}}Identify, understand\\ and discuss the impacts.\end{tabular}           \\ \hline
\end{tabular}
\end{table}

\subsection{Search for Studies}

After defining the research questions, following the protocol, the keywords related to the study were defined to create the search string, including: \textit{software engineering}, \textit{education}, \textit{gamification}, \textit{secondary studies}. 
From this, the search string was defined. This string was composed of the keywords and their synonyms and by the logical operators AND and OR, used to relate the defined search terms.

The combined terms led to the following \textbf{search string}:
\vspace{5pt}

\noindent\textit{(gamification OR gamified OR gamifies OR gamify OR gamifying) 
AND (''software engineering'' OR ``software design'' OR ``software process'' OR ``software requirements'' OR ``software testing'' OR ``software risk'' OR ``software integration'' OR ``software construction'' OR ``software implementation'' OR ``software verification'' OR ``software validation'' OR ``software metrics'' OR ``software maintenance'' OR ``software configuration'' OR ``project planning'' OR ``project assessment'') 
AND (education OR educational OR course OR teaching OR learning OR training) 
AND (``systematic literature review'' OR ``systematic review'' OR ``research review'' OR ``research synthesis'' OR ``research integration'' OR ``systematic overview'' OR ``systematic research synthesis'' OR ``integrative research review'' OR ``integrative review'' OR ``mapping study'' OR ``scoping study'' OR ``systematic mapping'')}
\vspace{5pt}

The next step was the \textbf{definition of the databases} for searching publications. The criteria used to select such databases were: classical databases in Computer Science; databases that index well-known journals and conferences in the area of Software Engineering; and active databases. Considering the criteria, the selected databases were ACM Library, IEEE Xplore, Springer, Scopus, and Web of Science. 

\subsection{Selection of Studies}

Once the databases and the search string were defined, we defined the \textbf{inclusion and exclusion criteria for selecting the studies}. These criteria were defined based on the study's objective and the defined research questions. The \textbf{inclusion criteria} are:

\begin{itemize}[leftmargin=*]
\item (I1) Secondary studies that consider gamification in the context of SE education.
\item (I2) Studies that are capable of answering the research questions of the tertiary study.
\end{itemize}

The \textbf{exclusion criteria}:

\begin{itemize}[leftmargin=*]
\item (E1) Studies that do not meet the inclusion criteria.
\item (E2) Studies that deal with serious games or other approaches related to gamification, and not specifically with gamification.
\item (E3) Studies that do not follow protocols for systematic mapping or systematic reviews.
\item (E4) Studies that are not in English.
\item (E5) Studies made available without full text, calls for papers, or books.
\item (E6) Duplicate studies or earlier versions of already selected studies.
\end{itemize}

With the criteria defined, \textbf{we conducted the search}, in January 2023, with no time/year limitation. A total of 670 publications were returned from the searches, considering the 5 databases. After conducting the searches, the \textbf{Phase 2} of the study began, which consisted of \textbf{reading the title, abstract, and keywords} to verify the inclusion and exclusion criteria and to exclude publications that did not fit the purpose of the study. The analysis of the articles was carried out by three researchers. At the end of the process, they met to discuss and analyze the articles that generated any doubts, reaching a consensus on the publications that should be included or excluded from the study. At this stage, 655 publications that did not meet the established criteria were excluded. Thus, \textbf{15 articles proceeded to the next phase (Phase 3)}, for a full reading.

\subsection{Data Extraction from publications}

\textbf{Phase 3} of the study consisted of the \textbf{full reading of the 15 publications}, and data extraction. The reading was carried out by three researchers, separately, each reading the 15 accepted articles. \textbf{Considering the selection criteria, 5 articles were excluded}, which did not meet the inclusion criteria. The researchers reached a consensus on the articles that did not fit the scope of the study through meetings for discussion.
In Table \ref{tab:numerosTrabalhos} the databases are presented with their respective numbers of publications returned in the search process, and excluded in the first and second filtering, as well as the final number of publications included in this study.

\begin{table}[!ht]
\caption{Number of publications returned, excluded in the filte, and included in the final result by database.}
\label{tab:numerosTrabalhos}
\centering
\footnotesize
\begin{tabular}{l|c|cc|c}
\hline
\multicolumn{1}{c|}{\multirow{2}{*}{\textbf{Database}}} & \multirow{2}{*}{\textbf{Returned}} & \multicolumn{2}{c|}{\textbf{Number of eliminated}} & \multirow{2}{*}{\textbf{Final}} \\ \cline{3-4}
\multicolumn{1}{c|}{} &  & \multicolumn{1}{l}{\textbf{1st Filter}} & \multicolumn{1}{l|}{\textbf{2nd Filter}} &  \\ \hline
ACM Library & 131 & \multicolumn{1}{c|}{129} & 0 & 2 \\ \hline
IEEE Xplore & 7 & \multicolumn{1}{c|}{6} & 1 & 0 \\ \hline
Scopus & 26 & \multicolumn{1}{c|}{24} & 1 & 1 \\ \hline
Springer & 492 & \multicolumn{1}{c|}{487} & 0 & 5 \\ \hline
Web of Science & 14 & \multicolumn{1}{c|}{9} & 3 & 2 \\ \hline
\textbf{Total} & \textbf{670} & \multicolumn{1}{c|}{\textbf{655}} & \textbf{5} & \textbf{10} \\ \hline
\end{tabular}
\end{table}

\subsubsection{Data extraction}
The \textbf{data extraction from the final set of publications} (10 publications), was carried out with the help of the StArt tool, in which fields for extraction were defined according to the research questions, including: title, authors, year, type of study, venue, gamification strategy, area of SE knowledge, elements of gamification, and positive and negative impacts of gamification. Such information was extracted from the articles after full reading and analyzed in search of answering the research questions proposed in this study. In Table~\ref{tab:finaisTrabalhos} the included studies are presented, year of publication, database, type of study, and publication vehicle, and place of publication, respectively.

\begin{table*}[!t]
\caption{Studies included in the tertiary study.}
\label{tab:finaisTrabalhos}
\centering
\footnotesize
\begin{tabular}{cllll}
\hline
\textbf{Ref} & \multicolumn{1}{c}{\textbf{Year}} & \multicolumn{1}{c}{\textbf{Base}} & \multicolumn{1}{c}{\textbf{Type of Study}} & \multicolumn{1}{c}{\textbf{Place}} \\ \hline
\citet{alhammad2018} & 2018 & Web of Science & Mapping &  \textit{Journal of Systems and Software} \\ \hline
\citet{jesus2018} & 2018 & ACM & Mapping & \textit{\begin{tabular}[c]{@{}l@{}}Brazilian Symposium on Systematic and Automated Software Testing\end{tabular}} \\ \hline
\citet{indriasari2020} & 2020 & Springer & Review & \textit{\begin{tabular}[c]{@{}l@{}}Education and Information Technologies Journal\end{tabular}} \\ \hline
\citet{machuca2018} & 2018 & Springer & Review & \textit{\begin{tabular}[c]{@{}l@{}}International Conference on Software Process Improvement\end{tabular}} \\ \hline
\citet{darejeh2016} & 2016 & Scopus & Review & \textit{\begin{tabular}[c]{@{}l@{}}International Journal of Human-Computer Interaction\end{tabular}} \\ \hline
\citet{dichev2017} & 2017 & Springer & Review & \textit{\begin{tabular}[c]{@{}l@{}}International journal of educational technology in higher education\end{tabular}} \\ \hline
\citet{toda2017} & 2017 & Springer & Mapping & \textit{\begin{tabular}[c]{@{}l@{}}Researcher Links Workshop: higher education for all\end{tabular}} \\ \hline
\citet{trinidad2018} & 2018 & Springer & Review & \textit{\begin{tabular}[c]{@{}l@{}}International Conference on Software Process Improvement and Capability Determination\end{tabular}} \\ \hline
\citet{mauricio2018} & 2018 & Web of Science & Mapping & \textit{\begin{tabular}[c]{@{}l@{}}Information and Software Technology Journal\end{tabular}} \\ \hline
\citet{souza2017} & 2017 & ACM & Mapping & \textit{\begin{tabular}[c]{@{}l@{}}International Conference on Software Engineering (SEET)\end{tabular}} \\  \hline

\end{tabular}
\end{table*}

\section{Results}
\label{sec:results}

The searches returned 670 publications that were present in at least one of the five selected databases. After applying inclusion and exclusion criteria, 10 secondary studies were included for extraction and analysis of results, comprising five systematic literature reviews and five systematic mappings. Half (5) of the studies included in the final selection were published in \textit{journals}, four (4) in conferences, and one was published in a \textit{workshop}. It is interesting to observe that there is no more than one study per venue, with all ten articles published in different venues.

Only six of these studies have a specific focus on investigating the use of gamification in SE education \cite{trinidad2018, souza2017, mauricio2018, machuca2018, alhammad2018, jesus2018}. Therefore, the decision was made to explore studies that provided analysis of primary studies on the use of gamification in computing education, as long as SE was part of the scope of the review results. Consequently, some selected studies~\cite{indriasari2020, darejeh2016, dichev2017, toda2017} do not address only SE studies but include primary studies from the area. Table \ref{tab:reqtrabalhos} presents the studies that could be used to answer each research question. It is important to highlight that the decision to use the studies for each research question is related to the scope of each secondary study analyzed. While some studies address all the questions covered by the tertiary study, the scope of others does not allow for certain research questions to be answered.

\begin{table}[!htb]
\caption{Relationship between QPs and selected studies.}
\label{tab:reqtrabalhos}
\centering
\footnotesize
\begin{tabular}{c|c}
\hline
\textbf{Research Question} & \textbf{Studies}    \\ \hline
RQ1 & \cite{alhammad2018, jesus2018, indriasari2020, darejeh2016, dichev2017, toda2017, trinidad2018, mauricio2018, souza2017} \\ \hline
RQ2 & \cite{alhammad2018, jesus2018, indriasari2020, machuca2018, dichev2017, toda2017, trinidad2018, mauricio2018, souza2017} \\ \hline
RQ3 & \cite{alhammad2018, machuca2018, dichev2017, mauricio2018} \\ \hline
RQ4 & \cite{alhammad2018,jesus2018,indriasari2020,darejeh2016,dichev2017, toda2017}    \\ \hline
\end{tabular}
\end{table}

Gamification in SE education is a relatively recent topic, thus the studies found date between the years 2016 and 2020. The highest number of secondary study publications related to the use of gamification in SE education was in 2018, with five publications. In 2016 there was one publication, three in 2017, and only one in 2020.

\subsection{RQ1 - Strategies used for the application of gamification in SE education?}

The first research question focuses on distinguishing how the studies address the use of gamification between the two types of strategies \cite{kapp2013gamification}. The difference between the two lies in the focus of game-originated elements being used in the structure or content, indicating whether the gamification is \textbf{structural} or \textbf{content}, as per \citet{kapp2013gamification}. As observed, the structure is understood as the medium of learning, being either a virtual or physical environment, while the content itself remains unchanged. Meanwhile, content gamification modifies the material by creating narratives to engage students, without necessarily having a gamified learning medium \cite{kapp2013gamification}. 

In summary, the mapping and review studies mostly address gamification as primarily structural. No studies were identified that use only the content gamification approach, as can be seen in Table~\ref{tab:estrategiasV2}. Some studies do not distinguish between structural and content approaches, seeking to differentiate gamification implementations by other more abstract aspects. In the studies by \citet{souza2017}, \citet{mauricio2018}, and \citet{darejeh2016}, both structural and content gamification are presented, with the authors noting that the structural approach is more frequently found due to its possession of more popular game elements.

\begin{table}[!t]
\caption{Relationship between study and strategy.}
\label{tab:estrategiasV2}
\centering
\footnotesize
\begin{tabular}{c|c}
\hline
\textbf{Strategy} & \textbf{Work} \\ \hline
Structural & \cite{jesus2018,indriasari2020,dichev2017,toda2017,trinidad2018} \\ \hline
Content & None \\ \hline
Structural and content & \cite{alhammad2018,darejeh2016,mauricio2018,souza2017} \\ \hline
Not applicable & \cite{machuca2018,mora2017} \\ \hline
\end{tabular}
\end{table}

In the study by \citet{trinidad2018}, the gamification approach used is structural gamification, with elements directed towards the learning environment, such as leaderboards and scoring. According to the authors, gamification occurs through four stages: business and requirements modeling, design, implementation, and monitoring. Moreover, the authors note that specific knowledge about the tools used is necessary to ensure proper gamification.

The systematic mapping by \citet{mauricio2018} presents both strategies, structural and content gamification. The authors provide a view of the use of game-derived elements related to the strategies of structural and content gamification. Thus, it is possible to understand how game elements operate in each gamification approach.

\citet{mauricio2018} mention the use of student scoring, medals for activities, and leaderboards to promote competition and motivate students. These elements are being used in the learning medium, which classifies this gamification as structural, according to \citet{mauricio2018}. With the participation of the same authors \cite{mauricio2018}, the systematic mapping by \citet{souza2017} also addresses both structural and content gamification.

In the systematic mapping by \citet{alhammad2018}, four ways to implement gamification are presented: the use or development of \textit{plug-ins}, the use of a generic gamified learning environment or the development of a proprietary gamified environment, which was the most identified method by the researchers \cite{alhammad2018}. Manual use of gamification in a non-virtual environment was also presented, incorporating game elements into the classroom. According to the authors, content gamification was more highlighted among the studies \cite{alhammad2018}. The analysis of the use of game elements indicates, according to the authors, that a significant majority do not follow any formal or structured process to gamify activities \cite{alhammad2018}.

In the review by \citet{jesus2018}, the gamification strategy used is structural, applying the concepts and elements of games in learning environments and in the execution of knowledge consolidation activities. The study materials used by the students were not gamified, as in cases that use narrative for content gamification. It can be observed in this study that structural gamification is more prevalent given the most frequent elements, such as leaderboards, progress levels, and scoring.

The Systematic Literature Review conducted by \citet{indriasari2020} focuses on structural gamification used in the student learning environment. The researchers focused on gamification in the classroom with peer-review activities. Such activities helped students participate and made the experience more engaging by conducting knowledge retention activities.

The study by \citet{darejeh2016} mentions both types of gamification and explains each separately. The authors identify elements aimed at structural gamification such as points, badges, and achievements. Meanwhile, they discuss that for content gamification it is necessary to gamify the material and work, for example using narrative to guide readings and knowledge consolidation exercises. They also pointed out that structural gamification is more frequently found, leaving content gamification participating in only 30\% of the primary studies.

\citet{dichev2017} study some cases of structural gamification. Most of the articles analyzed deal with gamified online courses, having gamified learning platforms and environments. According to the authors, most gamified courses are aimed at Computer Science and Information Technology. Thus, the gamification approach of the studies is focused on gamifying the learning structure that supports students in their motivation to perform activities.

Finally, the mapping by \citet{toda2017} highlights the negative effects of gamification in virtual environments. In this study, the data do not direct a response about the use of gamification, not detailing whether a structural or content approach is used. However, among the studies of the MSL, the elements most found by researchers may be directly related to structural gamification, such as the use of leaderboards, points, and levels in learning environments.

The studies of \citet{machuca2018} and \citet{mora2017} do not help answering RQ1. The mapping by \citet{machuca2018} focuses on gamification in software projects and management processes. The review by \citet{mora2017} addresses the \textit{design} of the \textit{frameworks}, emphasizing the stages of gamification. The authors do not present the relationship between the elements extracted from games and their use in the learning environment or content material, which is necessary to identify content and structural gamification.

\subsection{RQ2 - Which knowledge areas in SE have been gamified?}

Table \ref{tab:areas-eng2} presents the occurrences of each SE area in the studies. The SE areas follow the knowledge areas of the SWEBOK~\cite{10.5555/2616205}. It can be observed that Software Testing and Software Quality were the most studied areas, and Management (both SE and Configuration) and Maintenance appear in only one study each.

\begin{table}[!t]
  \caption{Gamified areas of Software Engineering.}
    \label{tab:areas-eng2}
\begin{center}
    \centering
    \footnotesize
    \begin{tabular}{ lcccccc }
        \hline
        {\textbf{Engineering areas}} & {\textbf{\cite{alhammad2018}}} & {\textbf{\cite{jesus2018}}} & {\textbf{\cite{machuca2018}}} & {\textbf{\cite{trinidad2018}}} & {\textbf{\cite{mauricio2018}}} & {\textbf{\cite{souza2017}}} \\ \hline
        {Software Testing} & \checkmark & \checkmark & & & \checkmark & \checkmark \\ \hline
        {Software Quality} & \checkmark & & \checkmark & & \checkmark & \checkmark \\ \hline
        {\textit{Software Design}} & \checkmark & & & & \checkmark & \checkmark \\ \hline
        SE Processes & \checkmark & & & & \checkmark & \checkmark\\ \hline
        {Software Requirements} & & & & \checkmark & \checkmark & \checkmark \\ \hline
        \multirow{1}{*} {SE Models and Methods } & & & & \multirow{1}{1em}{\checkmark} & \multirow{2}{1em}{\checkmark} & \multirow{1}{1em}{\checkmark} \\ \hline
        \multirow{1}{*} {SE Professional Practices} & & & & & \multirow{1}{1em}{\checkmark} & \multirow{1}{1em}{\checkmark}\\ \hline
        {Software Construction} & \checkmark & & \checkmark & & & \\ \hline
        {Software Maintenance} & \checkmark & & & & & \\ \hline
        \multirow{2}{12em} {Software Configuration Management} & \multirow{2}{1em}{\checkmark} & & & & & \\  \\ \hline
        \multirow{1}{12em} {SE Management} & & & \multirow{1}{1em}{\checkmark} & & &  \\ \hline
    \end{tabular}
    \vspace{-3mm}
\end{center}
\end{table}

In addition to the areas listed in Table~\ref{tab:areas-eng2}, the studies of \citet{toda2017}, \citet{indriasari2020}, \citet{darejeh2016}, and \citet{dichev2017} identified areas of Computing Foundations, Mathematical Foundations, and Engineering Foundations. These studies are not specifically from SE but include studies from this area in the results.

\subsection{RQ3 - What elements of gamification are used?}

\begin{figure}[htp]
\centering\includegraphics[width=\linewidth]{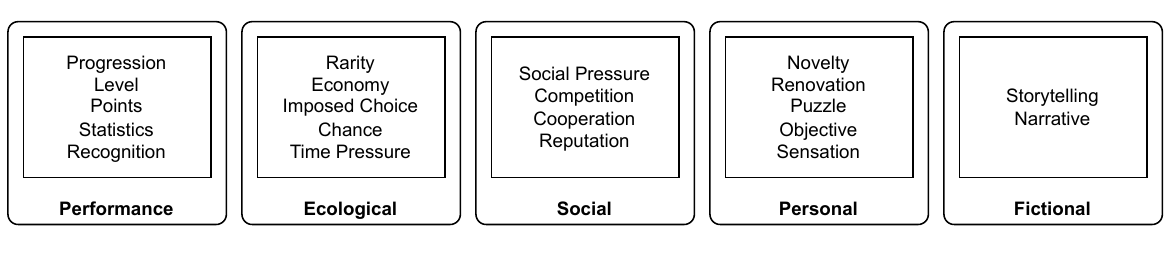}
\vspace{-3mm}
\caption{Gamification elements and dimensions from \cite{toda2019}.}
\vspace{-3mm}
\label{fig:gami-taxo}
\end{figure}

The gamification elements identified in the studies were categorized using the taxonomy proposed by \citet{toda2019}, which is a detailed taxonomy and can be used to design and evaluate gamification design in learning environments. An important feature of this taxonomy is its 5 dimensions, as they express the distinction between different types of \textit{feedback} and player interactions. Figure \ref{fig:gami-taxo} presents the 21 gamification elements across the 5 dimensions as described in the taxonomy of \citet{toda2019}.

\begin{table}[htp]
 \caption{Description of gamification dimensions.}
  \label{tab:dimensoes}
  
  \centering
  \resizebox{0.47\textwidth}{!}{
  \begin{tabular}{cc}
  \hline
  \textbf{Dimensions}  & \multicolumn{1}{c}{{\textbf{Description}}} \\ \hline
  \multirow{2}{*}{Performance} & Are elements related to the environment's response,  \\
  &  which can be used to provide feedback.  \\ \hline
  \multirow{3}{*}{Ecological} & This context is related to the environment   \\
  & where gamification is implemented, \\
  & the elements can be represented as properties.
  \\ \hline
  \multirow{2}{*}{Social}  & Related to interactions between players \\
  & presented in the environment. \\ \hline
  Personal  & Related to the player using the environment.  \\ \hline
  \multirow{3}{*}{Fictional} & Related to the user (through Narrative) and \\
  & the environment (through \textit{Storytelling}), linking \\
  & their experience in the context. \\ \hline
  \end{tabular}
  }
 
\end{table}

\begin{table*}[hb]
  \caption{Dimension, gamification elements and their descriptions.}
  \label{tab:desc-gami}
\centering
\footnotesize
\begin{tabular}{lll}
  \hline
  \multicolumn{1}{c}{\textbf{Dimension}} & \multicolumn{1}{c}{\textbf{Elements}} & \multicolumn{1}{c}{\textbf{Description}}   \\ \hline
  \multicolumn{1}{c}{\multirow{5}{*}{Performance}} &{Progression} & Provides extrinsic guidance to users about their progress in the environment, allowing them to locate themselves.  \\ \cline{2-3} 
  \multicolumn{1}{c}{} & \multirow{1}{*}{Level}  & Related to an extrinsic hierarchical layer that provides the user with advantages as they progress in the environment.   \\ \cline{2-3} 
  \multicolumn{1}{c}{} & \multirow{1}{*}{Point}  & A simple way to provide extrinsic feedback on user actions.   \\ \cline{2-3} 
  \multicolumn{1}{c}{} & \multirow{1}{*}{Statistics} & Related to the visual information provided by the environment to the learner (extrinsic).   \\ \cline{2-3} 
  \multicolumn{1}{c}{} & \multirow{1}{*}{Acknowledgement} & A type of extrinsic feedback that praises a specific set of player actions.  \\ \hline
  \multirow{5}{*}{Ecological}  & \multirow{1}{*}{Rarity} & Related to extrinsically limited resources in the environment that stimulate learners through a specific goal. \\ \cline{2-3} 
  & \multirow{1}{*}{Economy} & This concept is extrinsically related to any transactions that may occur in the environment.  \\ \cline{2-3} 
  & \multirow{1}{*}{Imposed Choice}  & This extrinsic concept occurs when the player faces an explicit decision they must make to advance in the environment.  \\ \cline{2-3} 
  & \multirow{1}{*}{Chance}   & This intrinsic concept is related to the random property of a particular event or outcome.  \\ \cline{2-3} 
  & \multirow{1}{*}{Time Pressure} & Related to the actual time used to pressure learners' actions (extrinsic).   \\ \hline
\multirow{4}{*}{Social}  & \multirow{1}{*}{Reputation}  & Related to titles that the learner can earn and accumulate within the environment, representing social status (intrinsic).  \\ \cline{2-3} 
   & \multirow{1}{*}{Cooperation} & Also an intrinsic concept (related to a task) where users must collaborate to achieve a common goal.  \\ \cline{2-3} 
   & \multirow{1}{*}{Competition} & An intrinsic concept, linked to a challenge where the user competes against another user to achieve a common goal.  \\ \cline{2-3} 
   & \multirow{1}{*}{Social Pressure} & This intrinsic concept is related to the social interactions that exert pressure on the learner.   \\ \hline
  \multirow{5}{*}{Personal}  & \multirow{1}{*}{Sensation} & Related to using the learners' senses to enhance the experience (intrinsic).  \\ \cline{2-3} 
   & \multirow{1}{*}{Objective} & Provides the player an end, or a purpose to perform the required tasks (intrinsic).  \\ \cline{2-3} 
   & \multirow{1}{*}{Puzzle}  & Related to learning activities in the environment.  \\ \cline{2-3} 
   & \multirow{1}{*}{Renewal}  & This concept is intrinsically related to redoing a task, event, or any of the types. \\ \cline{2-3} 
   & \multirow{1}{*}{Novelty} & Related to updates that occur within the environment (addition of new information, content, or game elements).  \\ \hline
  \multirow{2}{*}{Fictional}  & \multirow{1}{*}{Narrative}  & This intrinsic concept is the order of events as they occur in the game, through the user's experience.  \\ \cline{2-3} 
   & \multirow{1}{*}{\textit{Storytelling}} & It is how the story of the environment is told. It is told through text, voice, or sensory resources.  \\ \hline

  \end{tabular}

\end{table*}

Table \ref{tab:dimensoes} presents a description of each dimension and Table~\ref{tab:desc-gami} presents the description of each gamification element in the taxonomy of \citet{toda2019}. 

To correlate the gamification elements with the knowledge areas of SE, the data from \citet{mauricio2018}, \citet{machuca2018}, \citet{alhammad2018}, \citet{jesus2018} were compiled. Identifying 91 articles, with 267 gamification elements distributed across 11 knowledge areas of the SWEBOK~\cite{10.5555/2616205}.

\begin{figure}[!htp]
\centering\includegraphics[width=\linewidth]{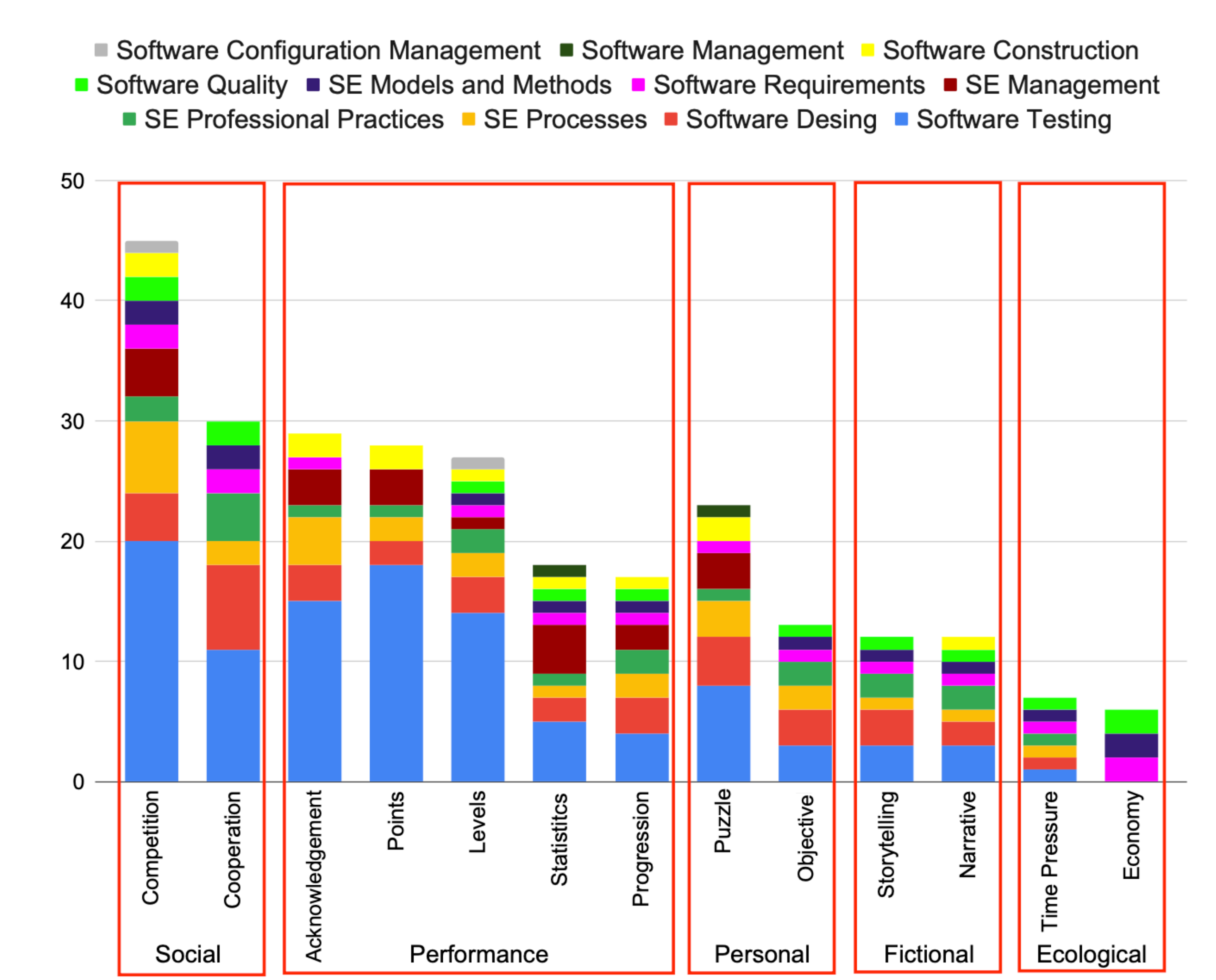}
\caption{Gamification elements and knowledge areas.}
\label{fig:elem-areas}
\end{figure}

Figure \ref{fig:elem-areas} presents the frequencies of gamification elements categorized according to the taxonomy of \citet{toda2019} classified by SE knowledge area~\cite{10.5555/2616205}. The most frequent dimension is Performance, with five different types of elements. The most frequent gamification element is Competition with 45 occurrences, followed by Cooperation with 30. Furthermore, it is observed that 13 (of the 21) different types of gamification elements have been employed in different SE knowledge areas.

The knowledge area of Software Testing shows the highest frequency with a value of 105. Meanwhile, the areas with the lowest frequency are Software Maintenance and Software Configuration Management, each with a frequency of 2. It is also observed that elements from all dimensions have been used in SE Methods and Models, SE Professional Practices, Requirements, Design, Testing, and Software Quality.

\subsection{RQ4 - What are the impacts of gamification on SE education?}

The goal of this research question was to understand the impacts of gamification on SE education, whether they are positive or negative. Of the 10 studies included in this tertiary study, only 5~\cite{alhammad2018,jesus2018,indriasari2020,darejeh2016,dichev2017} could address this question. Table~\ref{tab:positivos} presents a synthesis of the main positive impacts of gamification, the associated gamification elements, and the studies that cited them. Each positive aspect is related to a set of elements, i.e., these elements were used in the studies that reported the positive effects mentioned in the table.

\begin{table*}[!bh]
\caption{Main impacts of gamification.}
\label{tab:positivos}
  \centering
  \footnotesize
  \begin{tabular}{l|l|c|c|c|c|c|c|c|c|c|c|c|c|c|c|c|c}
  \hline
  \textbf{Effect} & 
  \textbf{Studies} & 
  \rotatebox{90}{\textbf{Progression}~~} & 
  \rotatebox{90}{\textbf{Level}~~} & 
  \rotatebox{90}{\textbf{Point}~~} & 
  \rotatebox{90}{\textbf{Statistics}~~} &
  \rotatebox{90}{\textbf{Acknowledgement}~~} & 
  \rotatebox{90}{\textbf{Narrative}~~} & 
  \rotatebox{90}{\textbf{\textit{Storytelling}}~~} & 
  \rotatebox{90}{\textbf{Objective}~~} & 
  \rotatebox{90}{\textbf{Puzzle}~~} & 
  \rotatebox{90}{\textbf{Cooperation}~~} & 
  \rotatebox{90}{\textbf{Competition}~~} & 
  \rotatebox{90}{\textbf{Rarity}~~} & 
  \rotatebox{90}{\textbf{Economy}~~} & 
  \rotatebox{90}{\textbf{Chance}~~} &
  \rotatebox{90}{\textbf{Time Pressure}~~} &
  \rotatebox{90}{\textbf{Reputation}~~} \\ \hline
  \hline
  \multicolumn{17}{c}{\textbf{Positive Effects}} & \\\hline \hline
  \begin{tabular}[l]{@{}l@{}}Engagement\end{tabular}  & \cite{alhammad2018,jesus2018,indriasari2020,darejeh2016,dichev2017} & + & ~  & + & + & + & + & + & + & + & + & + & + & + & + & + &  ~  \\ \hline
  Improve performance  & \cite{alhammad2018,jesus2018,indriasari2020,dichev2017} & + & + & + & + & + & + & + & + & + & + & + & + & + & + &  ~ \\ \hline
  Encourage good SE practices & \cite{alhammad2018,jesus2018} & + & + & + & + & + & + & + & + & + & + & + & + &  ~ &  ~ & ~\\ \hline
  \begin{tabular}[c]{@{}l@{}}Stimulate collaboration, leadership, and teamwork\end{tabular}& \cite{alhammad2018,jesus2018} & + & + & + & + & + & + & + & + & + & + & + & + &  ~ &  ~ &  ~  &  ~  \\ \hline
 \begin{tabular}[c]{@{}l@{}}Increase knowledge retention, and reduce attrition\end{tabular} & \cite{jesus2018,dichev2017} & ~ & + & + & + & + &  ~  &  ~  & + & + & + & + & ~ & + & + &  ~ &  ~  \\ \hline
  Improve performance  & \cite{alhammad2018,jesus2018,dichev2017} & + & + & + & + & + & + & + & + & + & + & + & + & + & + &  ~ &  ~  \\ \hline

\hline
  \multicolumn{17}{c}{\textbf{Negative Effects}} & \\\hline \hline
  \begin{tabular}[l]{@{}l@{}}Detrimental effect on motivation and satisfaction\end{tabular}  & \cite{dichev2017,toda2017} & 
  -- & -- & -- & -- & -- & -- & ~  & ~  & -- & -- & -- & ~  & -- & -- & ~ & --   \\ \hline
  Performance loss  & \cite{dichev2017,toda2017} & 
  -- & -- & -- & -- & -- & -- & ~  & ~  & -- & -- & -- & ~  & -- & -- & ~  & --  \\ \hline
  Undesirable behavior & \cite{toda2017} & 
  -- & -- & -- & -- & -- & -- & ~  & ~  & -- & -- & -- & ~  & -- & ~  & ~  & -- \\ \hline
  \begin{tabular}[c]{@{}l@{}}Indifference / lack of gamification influence\end{tabular} & \cite{toda2017} & 
  -- & -- & -- & -- & -- & -- & ~  & ~  & -- & -- & -- & ~  & -- & ~  & ~ & --  \\ \hline
 \begin{tabular}[c]{@{}l@{}}Decreasing effects / reduction in motivation\\ and engagement over time\end{tabular} & \cite{toda2017} & 
  -- & -- & -- & -- & -- & -- & ~  & ~  & -- & -- & -- & ~  & -- & ~  & ~ & --  \\ \hline
  \end{tabular}

\end{table*}

As a positive impact, the systematic mapping conducted by \citet{alhammad2018} grouped the benefits linked to gamification, presented in primary studies, into four categories: increased engagement, improved performance, encouragement to adopt best practices, and enhanced social skills.

The first category presented by \citet{alhammad2018} deals with the \textit{increase in engagement}, as gamified courses can stimulate and maintain students' interest in learning a particular topic. The mapping carried out by \citet{jesus2018} also provides some information that supports the findings of \citet{alhammad2018}, with the increase in student engagement being an important aspect of gamification. In the systematic reviews conducted by \citet{indriasari2020} and \citet{darejeh2016}, engagement was also one of the main positive aspects resulting from the use of gamification.

Another point mentioned by \citet{alhammad2018} was the \textit{improvement in student performance}. Gamification can enhance how students learn, making the learning process easier and more enjoyable, which can influence overall performance. The studies of \citet{indriasari2020} and \citet{jesus2018} also corroborate this category, highlighting the increase in performance in carrying out activities. The review by \citet{dichev2017} supports this information, citing benefits such as improved retention period and learning performance.

The third category pointed out by \citet{alhammad2018} was the \textit{encouragement to use good SE practices}, where gamification is not directly related to the learning process but rather to the application of the practices learned. The study by \citet{jesus2018} focused on gamification in teaching Software Testing. In this study, they found that gamification had a positive impact by encouraging developers to perform tests until they become a habit and improving training for conducting software tests or other software development activities.

The fourth category presented refers to the \textit{improvement of social and team skills}, with gamification playing a role in stimulating collaboration, leadership, and a sense of belonging. Again, the study of \citet{jesus2018} showed results that meet those of \citet{alhammad2018}, with the stimulation of collaboration and creativity being a positive aspect of gamification.

Another positive aspect found in the studies was the \textit{increase in knowledge retention, and the reduction of attrition}, cited by \citet{dichev2017} and \citet{jesus2018}. According to these studies, gamification can facilitate the process of content fixation, and consequently reduce student attrition in the learning process. The \textit{increase in motivation} was another point cited by \citet{dichev2017} and \citet{jesus2018}, and it also aligns with the findings of \citet{alhammad2018}.

Some negative points related to the adoption of gamification were also identified. The lower part of Table~\ref{tab:positivos} presents the synthesis of these main negative effects, the elements associated with the effects, and the studies that cited them.

The review by \citet{dichev2017}, in addition to bringing positive points, mentions some negative aspects of gamification, depending on the elements used. According to the findings of this RSL, the use of elements such as \textit{badges}, \textit{leaderboards}, and virtual coins can have a possible \textit{detrimental effect on motivation, satisfaction, and empowerment} of students. However, the authors also discuss that these results should be interpreted restrictively and may not apply to other contexts. Another study that brought up negative aspects was the mapping by \citet{toda2017}. \textit{Performance loss} was one of the most cited aspects in the primary studies identified by \citet{toda2017} (and echoed by \citet{dichev2017}), and for the authors, this problem arises from tasks and situations where gamification can hinder the learning process of students, due to demotivating effects. These effects are related to the following aspects: students did not understand the rules and this impaired their performance; students who were more active in the gamified activity scored lower than their peers because they were more focused on the gamified mechanics than on the evaluation; students felt that the gamified activities were too difficult, which also impacted their grades.

According to \citet{toda2017}, the second most common problem was \textit{undesirable behavior}. This occurred because gamification caused a different effect (positive or negative) in the learning context in which it was applied, either due to lack of or inadequate planning. Another negative aspect cited was indifference, which occurs when gamification does not influence, the students using the application. In general, according to the survey conducted by \citet{toda2017}, gamification in some contexts was \textit{indifferent}---it did not improve students' knowledge gain compared to the traditional learning method. Finally, \citet{toda2017} also report \textit{decreasing effects related to the loss of motivation and engagement} due to the implemented gamification process. Although similar to performance loss, considering that students' progress is hindered in both scenarios, they differ in that students' motivation and engagement decreased over time, which can also lead to performance loss.

\section{Discussions}
\label{sec:discussions}

When analyzing the studies---from the perspectives of structural and content gamification---it was noted that some studies do not distinguish between the two formats, treating gamification in a general sense. However, it can be considered that the two ways of using gamification can produce different results, as they can be applied in distinct contexts. It is important to highlight that the two types are not mutually exclusive and can exist simultaneously. They are more effective when combined \cite{kapp2013gamification}.

Additionally, the most gamified areas in SE were Software Testing and Software Quality. Through the tertiary study, we could not identify why these areas are the most targeted for gamified activities. Investigating the factors that lead teachers to choose the topic for using gamification could yield interesting results and assist other teachers in choosing content to work with using gamification.

The performance dimension has the highest frequency of elements, with 119 elements accounted for, where Acknowledgement, points, and levels are the highlights of this dimension with respective frequencies of 29, 28, and 27. However, the social dimension presents the elements with the highest frequency, notably competition and cooperation with respective frequencies of 45 and 30. From this information, it can be inferred that most often the elements responsible for providing \textit{feedback} to users are used to implement a logic of competition among users. Despite cooperation having the second highest frequency, competition shows significant popularity. On the other hand, several other elements have been little explored and have the potential to promote new research. Elements such as narrative, \textit{storytelling}, economy, and time pressure have the lowest frequencies, despite being common elements in the world of \textit{games}. Although some of these elements may not appear given the context of educational games, it would be important to investigate some of them, due to their importance in engagement, motivation, and explaining real problems in SE education.

Gamification in SE education can bring various benefits, such as increased motivation and engagement, improved performance and knowledge retention, and stimulation of collaboration, teamwork, and leadership. However, according to the analyzed articles, gamification can have the opposite effect depending on how it is used. Engagement, motivation, and increased performance, which are strong points in the use of this approach, can also be one of the biggest disadvantages, as noted in secondary studies.

Through the set of information collected, along with the analyses performed, it was possible to identify that gamification in SE education, although still little explored in theoretical studies, has a significant rate of utilization. One of the main contributions of this study relates to the positive and negative aspects of the use of gamification in SE education and the relationship of these aspects with the identified gamification elements. Aggregating such results in a single study is useful for teachers interested in adopting some aspect of gamification in their classes, assisting them in deciding which elements to adopt or avoid, and considering what they aim to achieve with the use of gamification in the educational environment. The results so far can guide towards the existing gaps in gamification and its process, allowing for an understanding of what is commonly used to break the standard and generate innovations.

\section{Threats to Validity}
\label{sec:threats}

This study adopted a robust protocol~\cite{kitchenham2004} to mitigate threats to validity. Three researchers conducted all phases, obtaining consistent results after resolving doubts in group meetings.

One threat was the limited number of articles to answer the RQs, leading to discussions based on a reduced number of them. Additionally, some studies addressed gamification in SE in a non-specific manner. To bring transparency, the text highlighted the articles relevant to each question.

Another threat was the scope of the included secondary studies, which could lead to a bias towards specific areas. For example, a specific article from the software testing area~\cite{jesus2018} might have influenced the results on gamification in testing and software quality.

\section{Conclusion}
\label{sec:conclusion}

Gamification has been used by various educators to improve the teaching and learning process. Accordingly, this study aimed to investigate the use of gamification in teaching Software Engineering (SE) courses. For this purpose, a tertiary study was conducted, as there were already recent secondary studies addressing this topic.

The results show that gamification has been used in its two formats (content and structural), both together and separately. However, structural gamification is used in the majority of the studies. Furthermore, it is possible to assert that one of the main benefits generated by gamification in the analyzed studies is the engagement and motivation of students. Gamification can also directly impact performance improvement, encouragement of good practices, and stimulation of collaboration, teamwork, and leadership. Regarding negative aspects, if gamification is not planned and applied correctly, it can cause demotivating effects and a loss of student performance.

Regarding the most gamified areas in SE, these are Software Testing and Software Quality. The most used gamification elements are competition and cooperation, both from the social dimension, according to the taxonomy of \citet{toda2019}.

This study provides relevant information on the use of gamification in Software Engineering (SE) education, applicable to teachers and researchers interested in the topic. It concludes that gamification can be an effective approach to increasing student motivation and engagement.

For future work, we suggest analyzing the context in which gamification is applied, considering variables such as environment (face-to-face or remote) and modality (synchronous or asynchronous). Furthermore, investigating the impacts of gamification in different areas of SE, considering the type of gamification used (structural or content) and its effects on educational outcomes, are promising lines of research. Exploring these issues can offer insights for developing more effective gamification strategies in SE education, optimizing its benefits, and addressing challenges.

\section*{Artifact Availability}
\label{sec:artifact}
The artifact used in this study is available in an online repository~\cite{repositorio}, including the spreadsheets and a .start file (used by the StArt tool).

\section*{Paper Availability}
This is the English Version of the paper published at the XXXVII Brazilian Symposium on Software Engineering (SBES 2023)~\cite{tonhao2023gamification}. The original paper is in Portuguese and is available at \url{https://dl.acm.org/doi/10.1145/3613372.3614193}. 

\noindent Tonhão, Simone; Shigenaga, Marcelo; Herculani, Julio; Medeiros, Andressa; Amaral, Aline; Silva, Williamson; Colanzi, Thelma; Steinmacher, Igor. 2023. Gamification in Software Engineering Education: a Tertiary Study. In XXXVII Brazilian Symposium on Software Engineering. 358–367.

\renewcommand{\acksname}{Acknowledgements}
\begin{acks}
The authors acknowledge the financial support from CAPES (Funding Code 001), CNPq (Processes 428994/2018-0 and 408812/2021-4), Fundação Araucária, and FAPERGS (ARD/ARC -- process 22/2551-0000606-0). Dr. Steinmacher's work is partially supported by the National Science Foundation (grants 230304 and 2247929).
\end{acks}

\bibliographystyle{ACM-Reference-Format}
\bibliography{_references.bib}

\end{document}